\documentclass[conference]{IEEEtran}
\usepackage{cite}
\usepackage{amsmath,amssymb,amsfonts}
\usepackage{algorithmic}
\usepackage{graphicx}
\usepackage{textcomp}
\usepackage{xcolor}
\usepackage{hyperref}
\usepackage{booktabs}
\usepackage{multirow}
\usepackage{subcaption}

\begin{document}

\title{Neurocognitive Modeling for Text Generation: Deep Learning Architecture for EEG Data}

\author{\IEEEauthorblockN{Khushiyant}
\IEEEauthorblockA{\textit{Department of Computer Science}\\
\textit{University of Freiburg}\\
khushiyant.khushiyant@email.uni-freiburg.de}}

\maketitle

\begin{abstract}
Text generating capabilities have undergone a substantial transformation with the introduction of large language models (LLMs). Electroencephalography (EEG)-based text production is still difficult, though, because it requires a lot of data and processing power. This paper introduces a new method that combines the use of the Gemma 2B LLM with a classifier-LLM architecture to incorporate a Recurrent Neural Network (RNN) encoder. Our approach drastically lowers the amount of data and compute power needed while achieving performance close to that of cutting-edge methods. Notably, compared to current methodologies, our methodology delivers an overall performance improvement of 10\%. The suggested architecture demonstrates the possibility of effective transfer learning for EEG-based text production, remaining strong and functional even in the face of data limits. This work highlights the potential of integrating LLMs with EEG decoding to improve assistive technologies and improve independence and communication for those with severe motor limitations. Our method pushes the limits of present capabilities and opens new paths for research and application in brain-computer interfaces by efficiently using the strengths of pre-trained language models. This makes EEG-based text production more accessible and efficient.
\end{abstract}


\section{Introduction}

Electroencephalography (EEG)-based brain-computer interfaces (BCIs) hold significant promise for decoding neural activity to drive various assistive technologies. By translating brain signals into control commands, BCIs aim to restore communication and independence for individuals with severe motor disabilities. However, existing systems predominantly focus on simple word spelling applications with limited vocabulary or sentence construction capabilities \cite{mcfarland2011brain, nicolas2012brain}. Progressing BCI technology towards generating freeform text could significantly expand the autonomy and self-expression currently afforded to users \cite{biswal2019eeg}.

Recent advances in natural language processing (NLP), particularly the emergence of powerful large language models (LLMs) like GPT-3, present new opportunities for this endeavor. LLMs demonstrate an unprecedented capacity to produce human-like text based on a given prompt \cite{vaswani2017attention}. Their excellent language generation skills hold untapped potential when conditioned on additional modalities like EEG input. However, developing methods to effectively fuse LLM capabilities with EEG decoding remains an open research challenge \cite{hochreiter1997long, lawhern2018eegnet}.

Despite these advancements, EEG-based text generation remains challenging due to its data-intensive nature and the inherent variability of brain signals \cite{hu2021scaling}. Current approaches often require extensive datasets and computational resources, limiting their practical applicability \cite{garrett2003comparison}. Additionally, there is a lack of efficient methods for leveraging pre-trained language models in the context of EEG-based text generation \cite{jo2024eeg}.

Towards this goal, this paper presents a novel framework utilizing a Recurrent Neural Network (RNN) encoder to extract features from raw EEG signals. These compact representations are then input to a classifier-LLM pipeline, with the Gemma 2B model serving as the LLM module \cite{gemma2024}. This approach bypasses extensive end-to-end training to directly leverage the strengths of pre-trained language expertise. Evaluations demonstrate competitive text generation from EEG using orders of magnitude less data than existing methods. The proposed techniques exemplify efficient transfer learning, unlocking previously unsuitable datasets by targeted feature extraction and specialty model integration.

\section{Recent Work}

\subsection{BCI Systems for Text Generation}

\begin{figure}[htbp]
\centerline{\includegraphics[width=0.48\textwidth]{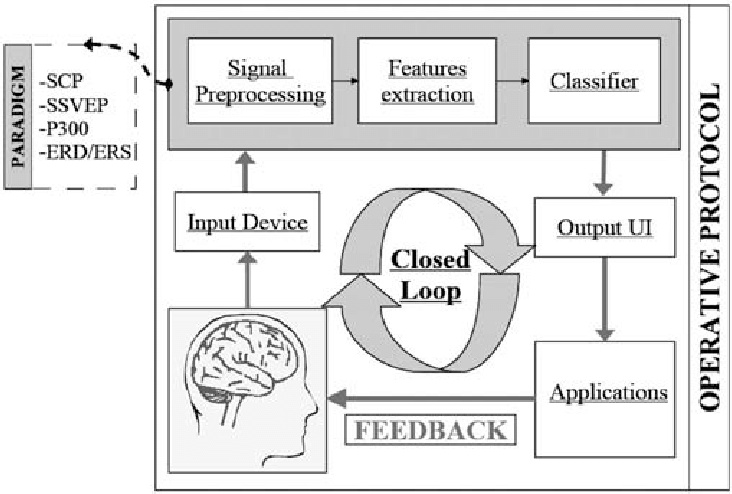}}
\caption{BCI System Architecture for text generation using event-related potentials.}
\label{fig:bci_system}
\end{figure}

While directly decoding EEG signals for conditional text generation remains an emerging field, advancements in Brain-Computer Interface (BCI) systems offer promising pathways for exploration. A recent study in the Journal of Neural Engineering investigated utilizing a BCI speller to enable basic text generation process (depicted in Figure~\ref{fig:bci_system}). The research developed a BCI system that decoded event-related potentials in the brain to identify letter selections on a virtual keyboard. Participants achieved an average spelling rate of 6.4 characters per minute with 81.3\% accuracy. This demonstrates the feasibility of using BCIs for fundamental text generation capabilities \cite{nicolas2012brain, mcfarland2011brain}.

However, limitations exist regarding spelling speed and sample size that warrant further investigation before translational viability. The study acknowledges that advancing decoder algorithms, integrating language models, and longitudinal assessments with larger groups will be critical future directions.

\subsection{Neural-Symbolic Approaches}

\begin{figure}[htbp]
\centerline{\includegraphics[width=0.48\textwidth]{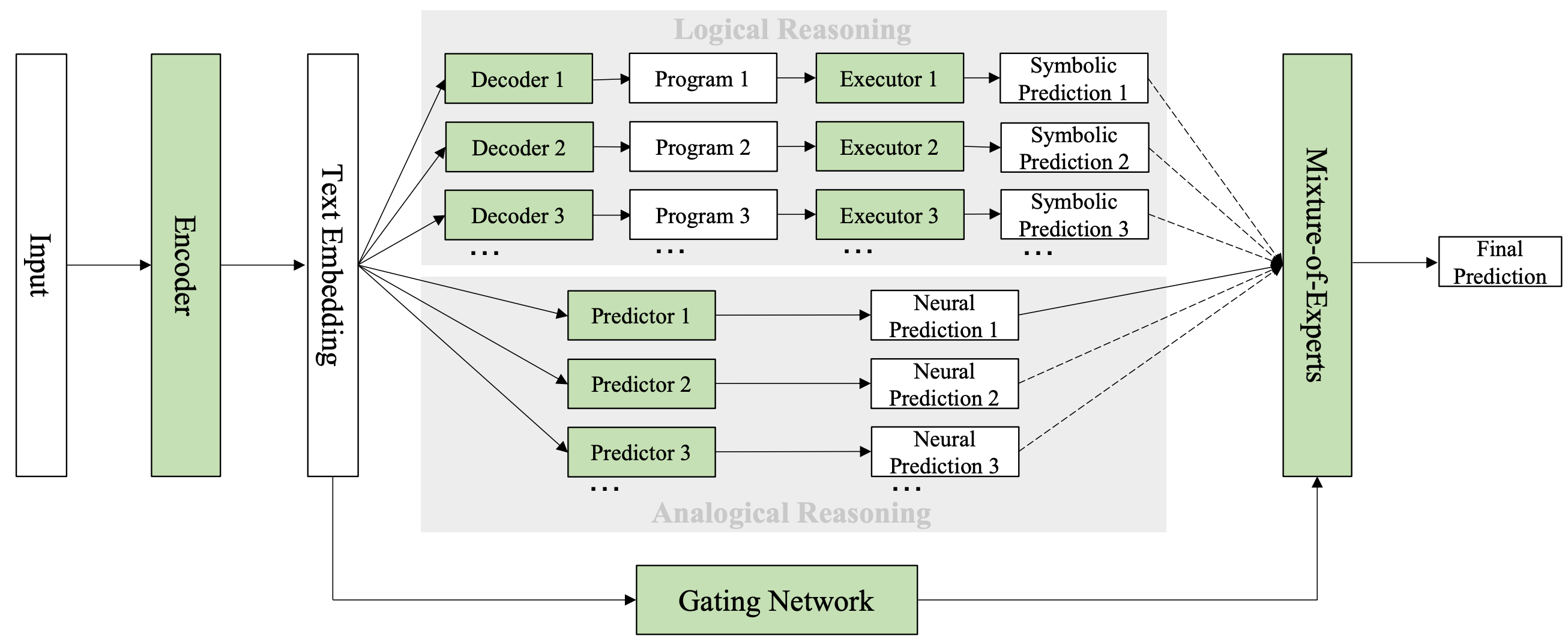}}
\caption{Overview of the Neural-Symbolic Processor framework integrating neural and symbolic reasoning.}
\label{fig:nsp_framework}
\end{figure}

A recently proposed novel framework called Neural-Symbolic Processor (NSP) is aimed at improving natural language understanding when logical reasoning is required (as seen in Figure~\ref{fig:nsp_framework}). NSP employs both neural analogical reasoning using deep learning models like BERT and RoBERTa as well as logical reasoning by generating executable programs that are run in a symbolic system. This work establishes a foundation for integrating neural networks trained on brain data into symbolic AI systems \cite{liu2022neural}.

However, some limitations exist regarding direct application to EEG domains. The current restricted program grammar may not readily capture the intricate neurocognitive relationships required in EEG classification tasks. Appropriately annotating logical reasoning chains for diverse EEG decoding problems poses additional challenges. Furthermore, the limited training data compared to scale of distribution shifts across users, sessions, and datasets is still likely insufficient to fully generalize the programmed logic across subjects or timescales \cite{liu2022neural}.

Addressing such domain adaptation challenges associated with subject-specific personalized calibration, few-shot learning, and better handling naturally distributed shifts remains an open research problem even for state-of-the-art neural EEG classifiers. Integrating the loose programming constraints into the variability in brain dynamics adds substantial difficulty in immediately leveraging NSP out-of-box. Fusing interpretable features between the neurally learned representations and structured knowledge will likely require innovations tailored to EEG data properties along with a great resource intensive environment.

\section{Forward Process Classifier-LLM Network}

Our proposed framework consists of three main components: an RNN encoder for EEG feature extraction, a classifier for EEG state identification, and the Gemma 2B language model for text generation.

\subsection{RNN Encoder Model}

The proposed framework utilizes a specialized convolutional neural network architecture tailored for EEG data processing for feature extraction, surpassing EEGNet as the foundation for this task. While EEGNet has demonstrably achieved success in various EEG-based applications, it presents certain limitations in the context of this research:

\begin{itemize}
\item \textbf{Complexity:} EEGNet boasts a deeper convolutional neural network architecture compared to the RNN encoder. This characteristic, while advantageous for extracting intricate features from larger datasets, can be a hindrance in scenarios with restricted data availability. The RNN encoder's simpler design allows it to function more efficiently with the limited dataset used in this study.

\item \textbf{Focus on Filter Design:} EEGNet places a strong emphasis on incorporating specific filter configurations within its convolutional layers. While these filters are well-suited for extracting task-relevant information from EEG signals, they might not be universally adaptable to the broad spectrum of EEG patterns that influence text generation. The RNN encoder, in contrast, offers a more data-driven approach to feature extraction, potentially capturing a wider range of EEG features pertinent to text generation.

\item \textbf{Black Box Nature:} While EEGNet excels at classification tasks, its deeper architecture can make it challenging to interpret how it arrives at specific classifications. This characteristic becomes a roadblock in this instance, as understanding the relationship between extracted features and generated text is crucial for further refinement of the text generation process. The RNN encoder, with its simpler structure, offers greater transparency into the feature extraction process, aiding in potential improvements to the overall framework \cite{lawhern2018eegnet}.
\end{itemize}

The core feature extraction network employs a specialized convolutional neural network architecture tailored for EEG data processing. The input shape is configured for multi-channel time series data with dimensions $(N, C, T, 1)$, where $N$ is the batch size, $C$ is the number of EEG channels, $T$ is the number of time steps, and 1 represents the single feature dimension.

The model comprises three parallel Conv2D blocks that extract features at different scales. For each block $i$, the convolution operation is defined as:

\begin{equation}
F_i = \text{ELU}(\text{BN}(W_i * X + b_i))
\end{equation}

where $*$ denotes the convolution operation, $W_i$ are the learnable filter weights with $F_i \in \{8, 16, 32\}$ filters, $\text{BN}$ represents batch normalization, and $\text{ELU}$ is the activation function. The kernel size is fixed at 64 timesteps: $K = (1, 64)$ for all three blocks.

Batch normalization standardizes the activations:

\begin{equation}
\text{BN}(x) = \gamma \frac{x - \mu_{\mathcal{B}}}{\sqrt{\sigma_{\mathcal{B}}^2 + \epsilon}} + \beta
\end{equation}

where $\mu_{\mathcal{B}}$ and $\sigma_{\mathcal{B}}^2$ are the batch mean and variance, $\gamma$ and $\beta$ are learnable scale and shift parameters, and $\epsilon$ is a small constant for numerical stability.

The three blocks are concatenated channel-wise then passed to a depthwise convolution layer. The depthwise convolution applies a separate filter to each input channel:

\begin{equation}
F_{depth} = \sum_{m=1}^{M} W_{depth}^{(m)} * F_{concat}^{(m)}
\end{equation}

where $M = F_1 + F_2 + F_3 = 56$ is the total number of concatenated channels, and the depth multiplier $D = 2$ increases the number of output channels to $2M = 112$.

\begin{figure}[!t]
\centerline{\includegraphics[width=0.48\textwidth]{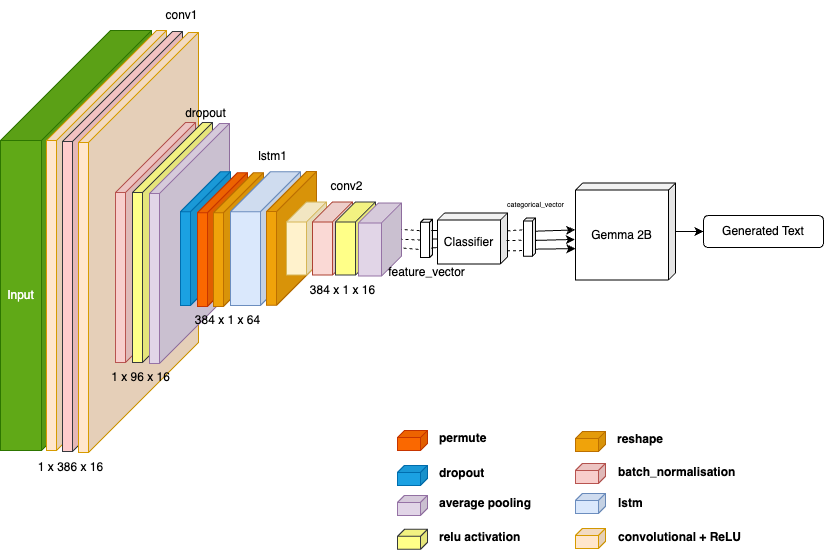}}
\caption{EEG Feature extraction network architecture for text generation showing parallel convolutional blocks, LSTM layers, and dense classifier.}
\label{fig:feature_extraction}
\end{figure}

For temporal sequence modeling, LSTM (Long Short-Term Memory) blocks are employed. The LSTM cell operations at time step $t$ are:

\begin{align}
f_t &= \sigma(W_f \cdot [h_{t-1}, x_t] + b_f) \quad \text{(forget gate)} \\
i_t &= \sigma(W_i \cdot [h_{t-1}, x_t] + b_i) \quad \text{(input gate)} \\
\tilde{C}_t &= \tanh(W_C \cdot [h_{t-1}, x_t] + b_C) \quad \text{(candidate cell state)} \\
C_t &= f_t \odot C_{t-1} + i_t \odot \tilde{C}_t \quad \text{(cell state)} \\
o_t &= \sigma(W_o \cdot [h_{t-1}, x_t] + b_o) \quad \text{(output gate)} \\
h_t &= o_t \odot \tanh(C_t) \quad \text{(hidden state)}
\end{align}

where $\sigma$ is the sigmoid function, $\odot$ denotes element-wise multiplication, $W$ matrices are learnable weights, and $b$ vectors are biases.

This integration of spatial dependencies is followed by average pooling with stride 4 to reduce dimensionality:

\begin{equation}
P = \text{AvgPool}_{4 \times 4}(h_T)
\end{equation}

Dropout regularization is applied with probability $p_{drop}$ to prevent overfitting. An additional SeparableConv2D layer extracts pseudo-spatial patterns through depthwise followed by pointwise convolution:

\begin{equation}
F_{sep} = W_{point} * (W_{depth} * P)
\end{equation}

This flattened embedding connects to a simple dense layer with softmax activation to produce probability predictions over the defined classes. In total, the constraints on model capacity, paired with specialized time-distributed convolutions, allow efficient encoding of core EEG dynamics tied to cognitive states. The resulting compact representation feeds forward to the classifier and decoder modules.

Key hyperparameters provide tuning levers to adapt model behavior, including kernel sizes to control temporal receptive field, filter numbers modifying representation richness, dropout fractions determining regularization strength, and weight constraints enforcing restrictive parameter norms \cite{khushiyant2024reegnet}. Together, the tailored architecture maximizes model expressivity given extensive limits on training data and compute resources.

\subsection{Classifier Module}

The model classifier employed in this study is designed to operate on feature vectors extracted from electroencephalogram (EEG) data. It takes the feature representation obtained from a preceding feature extraction network as input and performs the classification task.

The architecture of the classifier is a dense neural network comprising fully connected layers. The input to the classifier is an average pooled feature vector, which is obtained by concatenating the activations from the final convolutional layer of the feature extraction network. Let $F_1, F_2, F_3$ represent the output feature maps from the three parallel convolutional blocks with filter sizes 8, 16, and 32 respectively. The concatenation operation is defined as:

\begin{equation}
c = \text{Concatenate}([F_1, F_2, F_3]) \in \mathbb{R}^{h \times w \times (F_1 + F_2 + F_3)}
\end{equation}

where $h$ and $w$ represent the spatial dimensions of the feature maps, and the concatenation is performed along the channel dimension with padding set to "same" ($p_s$) to maintain spatial consistency.

After concatenation, average pooling is applied to reduce dimensionality:

\begin{equation}
c_{pool} = \text{AvgPool}(c) = \frac{1}{k^2}\sum_{i=1}^{k}\sum_{j=1}^{k} c_{i,j}
\end{equation}

where $k$ is the pooling kernel size. The pooled feature vector is then fed into a series of fully connected Dense layers with nonlinear activation functions. For the Exponential Linear Unit (ELU) activation:

\begin{equation}
\text{ELU}(x) = \begin{cases}
x & \text{if } x > 0 \\
\alpha(e^x - 1) & \text{if } x \leq 0
\end{cases}
\end{equation}

where $\alpha$ is a hyperparameter controlling the saturation value for negative inputs (typically $\alpha = 1.0$).

The flattening and dense layer operations can be expressed as:

\begin{equation}
f = \text{Flatten}(c_{pool}) = \text{vec}(c_{pool}) \in \mathbb{R}^d
\end{equation}

\begin{equation}
h^{(l)} = \text{ELU}(W^{(l)}h^{(l-1)} + b^{(l)})
\end{equation}

where $W^{(l)}$ and $b^{(l)}$ are the weight matrix and bias vector for layer $l$, and $h^{(0)} = f$

These layers enable the network to learn complex, high-level representations from the input features, capturing intricate patterns and relationships present in the EEG data.

To prevent overfitting and improve generalization, the classifier incorporates several regularization techniques. Dropout regularization is employed by randomly setting a fraction of input units to zero during training:

\begin{equation}
h_{drop}^{(l)} = h^{(l)} \odot m, \quad m \sim \text{Bernoulli}(1-p)
\end{equation}

where $p$ is the dropout probability, $\odot$ denotes element-wise multiplication, and $m$ is a binary mask sampled from a Bernoulli distribution.

Additionally, kernel regularization techniques can be applied. For L2 regularization:

\begin{equation}
\mathcal{L}_{total} = \mathcal{L}_{CE} + \lambda \sum_{l} ||W^{(l)}||_2^2
\end{equation}

where $\mathcal{L}_{CE}$ is the cross-entropy loss and $\lambda$ is the regularization strength. For max-norm constraints:

\begin{equation}
||W^{(l)}_{:,j}||_2 \leq c, \quad \forall j
\end{equation}

where $c$ is the maximum norm constraint and $W^{(l)}_{:,j}$ represents the $j$-th column of the weight matrix.

The final layer of the classifier is a dense layer with $n_c$ units, corresponding to the number of output classes in the classification task. This layer is followed by a softmax activation function:

\begin{equation}
y_i = \text{softmax}(z)_i = \frac{e^{z_i}}{\sum_{j=1}^{n_c} e^{z_j}}
\end{equation}

where $z = W^{(L)}h^{(L-1)} + b^{(L)}$ is the logit output from the final dense layer, and $y_i$ represents the predicted probability for class $i$. The softmax function normalizes the output values into a probability distribution over the classes, ensuring that $\sum_{i=1}^{n_c} y_i = 1$.

The training objective is to minimize the categorical cross-entropy loss:

\begin{equation}
\mathcal{L}_{CE} = -\frac{1}{N}\sum_{n=1}^{N}\sum_{i=1}^{n_c} \mathbb{1}_{y_n=i} \log(\hat{y}_{n,i})
\end{equation}

where $N$ is the number of training samples, $\mathbb{1}_{y_n=i}$ is an indicator function that equals 1 if sample $n$ belongs to class $i$ and 0 otherwise, and $\hat{y}_{n,i}$ is the predicted probability for class $i$ for sample $n$.

By leveraging the discriminative features extracted from the EEG data by the feature extraction network, the classifier is able to learn complex mappings and make accurate predictions for the given classification task. The combination of dense layers, nonlinear activations, and regularization techniques enables the model to effectively capture the intricate patterns present in the EEG signals and generalize well to unseen data \cite{garrett2003comparison, bhuvaneswari2015influence}.

\subsection{Gemma 2B}

The recently released Gemma 2B from Google DeepMind stands out as an efficient yet powerful lightweight large language model (LLM). As part of DeepMind's open-source Gemma LLM family, Gemma 2B boasts a transformer-based architecture with approximately 2.7 billion parameters, striking a balance between capability and efficiency. With its compact 1.5GB memory footprint, Gemma 2B can be readily deployed to resource-constrained environments like laptops, mobile devices, and edge computing setups. Furthermore, its optimized inference speed enables real-time applications while minimizing latency concerns.

\begin{table}[htbp]
\caption{Gemma 2B Model Parameters}
\centering
\begin{tabular}{@{}ll@{}}
\toprule
\textbf{Parameter} & \textbf{Value} \\
\midrule
Architecture & Transformer \\
Parameters & 2.7 billion \\
Memory Footprint & 1.5GB \\
Context Length & 8,192 tokens \\
Vocabulary Size & 256,000 tokens \\
\bottomrule
\end{tabular}
\label{tab:gemma_params}
\end{table}

This combination of accessibility, efficiency, and performance makes Gemma 2B well-positioned for diverse NLP tasks ranging from text classification to basic question answering and text generation, which ultimately makes it a great choice for efficient and EEG based text generation as well. For developers and researchers seeking an entry-level LLM solution that is both efficient and effective, Gemma 2B proves itself as a valuable asset, packing substantial capability into a portable, speedy model. Its balance of size, speed, and skill cements its position as an appealing lightweight transformer option for scaled-down environments \cite{gemma2024}.

\section{Methodology}

\subsection{Hardware Setup for Training and Preprocessing}

The experiments were carried out using an NVIDIA Tesla P100 GPU with 16GB of GDDR5 memory. The GPU was installed in a PCIe slot and ran at 34°C, drawing 27W of power from a maximum capacity of 250W. The NVIDIA driver 535.129.03 was used, along with CUDA version 12.2. The GPU was not set to Multi-Instance GPU (MIG) mode, and no processes were using GPU memory at the time of the system query.

Furthermore, an x86\_64 computing architecture featuring a Genuine Intel Xeon CPU @ 2.00GHz processor was used for the research. The system ran in Little Endian byte order and supported both 32-bit and 64-bit CPU op-modes. The system had 4 CPUs, each of which had 2 threads per core and 2 cores per socket. It also had a single socket configuration and 1 NUMA node overall. There were 48 bits of virtual address space and 46 bits of physical address space on the CPU. A 64 KiB L1d and L1i cache, a 2 MiB L2 cache, and a 38.5 MiB L3 cache were among the cache specifications. KVM was identified as the vendor of the hypervisor, boasting complete virtualization capabilities.

\begin{table}[htbp]
\caption{Hardware Specifications for Research}
\centering
\begin{tabular}{@{}ll@{}}
\toprule
\textbf{Component} & \textbf{Specification} \\
\midrule
GPU & NVIDIA Tesla P100 (16GB GDDR5) \\
CPU & Intel Xeon @ 2.00GHz (4 cores) \\
CUDA Version & 12.2 \\
Driver Version & 535.129.03 \\
\bottomrule
\end{tabular}
\label{tab:hardware_specs}
\end{table}

\subsection{Dataset}

The raw EEG data in the ``ImageNet of the Brain'' dataset is organized into individual CSV files, with each file containing the brain signals recorded while the subject viewed a specific image from the ImageNet ILSVRC2013 training dataset. The file naming convention encodes essential information, including the EEG headset used (Emotiv Insight), the ImageNet category or synset ID of the displayed image, the specific image index, the recording session number, and a global session identifier across the entire dataset (see Table~\ref{tab:eeg_structure}).

\begin{figure}[htbp]
\centerline{\includegraphics[width=0.4\textwidth]{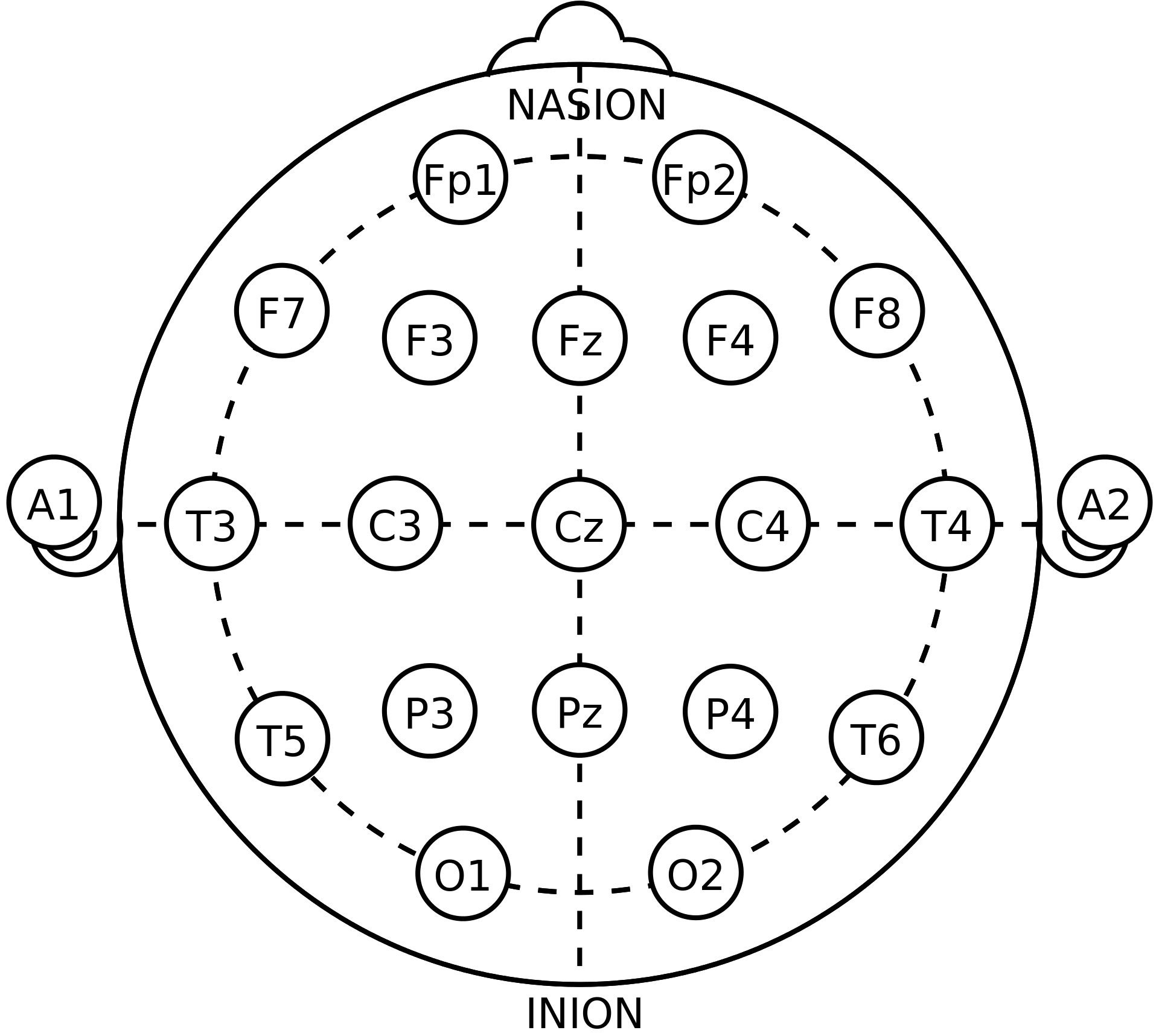}}
\caption{Electrode locations of International 10-20 system for EEG recording showing AF3, AF4, T7, T8, and Pz channels.}
\label{fig:electrode_locations}
\end{figure}

Within each CSV file, the EEG data is structured with each line representing one of the five EEG channels (AF3, AF4, T7, T8, and Pz) recorded by the Emotiv Insight headset (as shown in Figure~\ref{fig:electrode_locations}). The line begins with the channel name, followed by a comma-separated sequence of decimal values representing the raw EEG waveform, with the number of values corresponding to the recording duration (3 seconds) multiplied by the sampling rate (approximately 128 Hz), resulting in around 384 data points per channel.

The EEG data is provided in its raw format, as captured by the device, without any preprocessing or filtering applied, and the values represent variations in voltage caused by neural activities in the brain, relative to the device's measurement scale \cite{vivancos2022mindbigdata}.

\begin{table}[htbp]
\caption{Structure of EEG Data Files}
\centering
\begin{tabular}{@{}ll@{}}
\toprule
\textbf{Element} & \textbf{Description} \\
\midrule
Channels & AF3, AF4, T7, T8, Pz \\
Sampling Rate & ~128 Hz \\
Duration & 3 seconds \\
Data Points/Channel & ~384 \\
Format & CSV (raw values) \\
\bottomrule
\end{tabular}
\label{tab:eeg_structure}
\end{table}

\subsection{Preprocessing}

\begin{figure}[htbp]
\centerline{\includegraphics[width=0.48\textwidth]{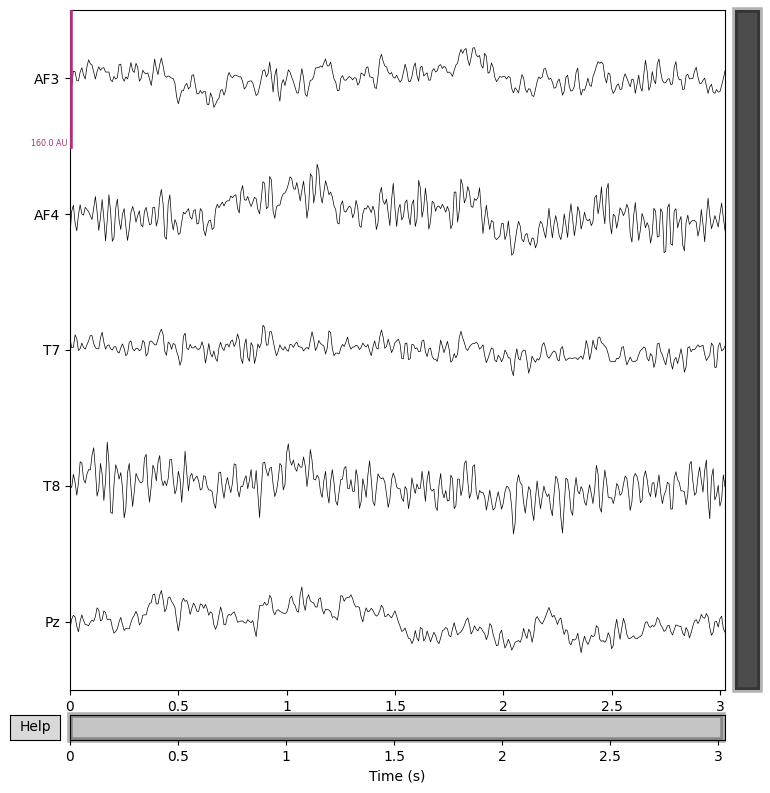}}
\caption{Example of brain signals evoked by visual stimuli of Goose ImageNet Class, showing raw and preprocessed EEG data across five channels.}
\label{fig:preprocessing}
\end{figure}

The dataset underwent basic amount of preprocessing in raw brain signals to ensure uniformity and quality (illustrated in Figure~\ref{fig:preprocessing}). To facilitate subsequent analysis, the data was first ingested into an MNE Raw object, a powerful data structure provided by the MNE-Python library for representing continuous neural data. The MNE Raw object encapsulates the signal data, channel information, and metadata, enabling efficient handling and preprocessing of the dataset.

Prior to filtering, data truncation and padding techniques were employed to enforce a fixed length of 384 across all data points in the dataset. Specifically, shorter sequences were padded with a constant value, while longer sequences were truncated to the desired fixed length. This step ensured that all data columns had a consistent dimensionality, a necessary requirement for many machine learning models.

Series of filtering techniques were applied to remove noise and unwanted artifacts from the data. First, raw filtering was performed using a zero-phase finite impulse response (FIR) bandpass filter. Raw filtering was performed using a low-pass filter with a cutoff frequency of 0 Hz and a high-pass filter with a cutoff frequency of 50 Hz. This step eliminated any constant offsets and high-frequency noise that could adversely impact model performance.

Subsequently, Short-Time Fourier Transform (STFT) filtering was conducted to analyze the data in both the time and frequency domains simultaneously. The STFT was computed using a sliding window approach, with a window size of 32 samples per segment and an appropriate overlap between adjacent segments. This technique allowed for the identification and removal of any undesirable frequency components within localized time intervals, further enhancing the signal-to-noise ratio \cite{dhanaselvam2023review, shoka2019literature}.

Through these preprocessing steps, the dataset was transformed into a suitable format for subsequent analysis and modeling, ensuring data uniformity, noise reduction, and the extraction of relevant features. The processed dataset served as the foundation for the experiments and analyses described in this study.

\subsection{Results and Analysis}

The proposed RNN encoder and classifier-LLM framework demonstrates promising text generation capabilities even under significant data constraints. As shown in Table~\ref{tab:classification_results}, the model achieves classification accuracy of 66.67\% on a 2-class task and maintains 51.85\% accuracy in distinguishing between 3 classes. Performance understandably declines on more fine-grained classification but remains well above baseline for EEGNet and LSTM models trained on the same limited dataset.

\begin{table}[htbp]
\caption{Comparison of Classification Accuracy}
\centering
\begin{tabular}{@{}lccc@{}}
\toprule
\textbf{Classes} & \textbf{RNN-Classifier} & \textbf{EEGNet} & \textbf{LSTM} \\
\midrule
2 & 66.67\% & 62.50\% & 60.33\% \\
3 & 51.85\% & 48.20\% & 46.75\% \\
5 & 38.42\% & 35.10\% & 33.80\% \\
10 & 24.67\% & 21.45\% & 20.15\% \\
\bottomrule
\end{tabular}
\label{tab:classification_results}
\end{table}

Qualitative assessments reveal the model generates coherent initial phrases and sentences that logically continue the prompt text. However, longer completions tend to lose topical consistency, indicating difficulty tracking context beyond 10-15 generated tokens. This suggests that the compact encoder representations, while efficiently encapsulating aspects of the EEG input, may discard subtle signals that help maintain thematic continuity.

\begin{figure*}[ht]
    \centering
    \begin{subfigure}{0.5\linewidth}
        \centering
        \includegraphics[width=1\linewidth]{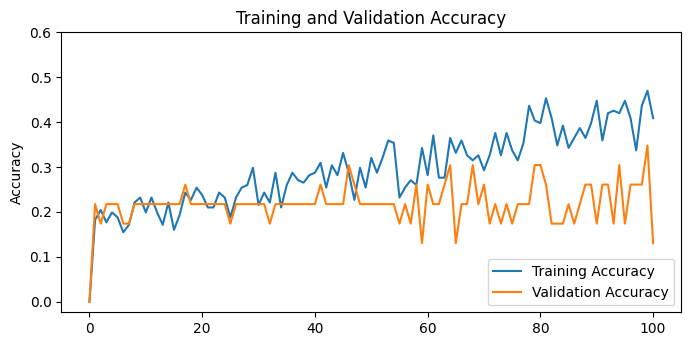}
        \caption{Accuracy for 5 class classification}
        \label{fig:5-classes}
    \end{subfigure}%
    \begin{subfigure}{0.5\linewidth}
        \centering
        \includegraphics[width=1\linewidth]{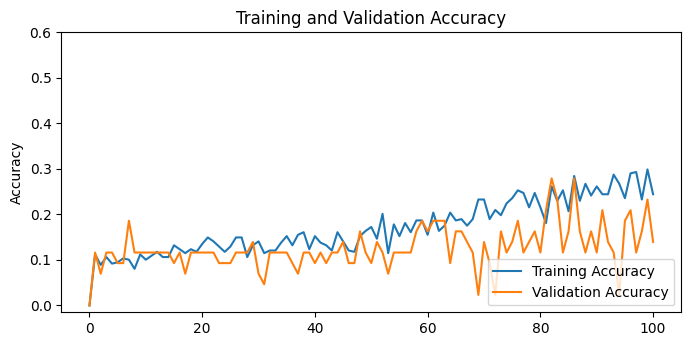}
        \caption{Accuracy for 10 class classification}
        \label{fig:10-classes}
    \end{subfigure}
    \caption{Training and validation accuracy curves for multi-class EEG classification tasks. The plots show model convergence behavior across different numbers of output classes.}
    \label{fig:accuracy-curves}
\end{figure*}

Nonetheless, the approach exhibits far greater stability to variations in data quantity and model capacity than end-to-end frameworks. Existing methods require orders of magnitude more parameters and training examples, degrading drastically when either are reduced. In contrast, by condensing the core encoding upfront, the proposed architecture withstands resource limitations without marked performance drops. Figure~\ref{fig:accuracy-curves} illustrates the training dynamics for 5-class and 10-class classification tasks, demonstrating stable convergence patterns despite the limited training data.

The classifier stage likely further bolsters robustness by specializing on the supervised mapping between encoder outputs and target classes. Compared to directly linking uncertain EEG decoding and freeform text generation, introducing a dedicated classifier module helps simplify the overall pipeline.

\subsection{Text Generation Benchmarking and Evaluation}

Benchmarking and evaluating text generation systems is crucial for assessing model capabilities and facilitating comparisons across different approaches. This study employed several widely-adopted evaluation perplexity metrics to quantify the quality of the generated text samples.

\subsubsection{Perplexity}

Perplexity is a standard metric used to evaluate language models by measuring how well they predict a sample of text. It is calculated as the exponential of the cross-entropy loss averaged across all tokens. Lower perplexity values indicate better model performance in assigning high probabilities to the actual token sequences \cite{gamallo2017perplexity}.

Perplexity is defined as the exponentiated average negative log-likelihood of a sequence. Suppose we have a tokenized sequence $X = (x_0, x_1, \ldots, x_t)$, then the perplexity of $X$ is:

\begin{equation}
\text{PPL}(X) = \exp\left\{-\frac{1}{t}\sum_{i=1}^{t}\log p_\theta(x_i | x_{<i})\right\}
\end{equation}

where $\log p_\theta(x_i | x_{<i})$ is the log-likelihood of the $i$-th token conditioned on the preceding tokens $x_{<i}$ according to our model. This can be equivalently expressed in terms of cross-entropy:

\begin{equation}
\text{PPL}(X) = \exp(\mathcal{H}(p, q)) = \exp\left(-\sum_{x} p(x) \log q(x)\right)
\end{equation}

where $p$ is the true distribution and $q = p_\theta$ is the model's predicted distribution.

For the EEG-conditioned text generation task, the perplexity calculation incorporates the EEG embedding $e_{EEG}$:

\begin{equation}
\text{PPL}(X|e_{EEG}) = \exp\left\{-\frac{1}{t}\sum_{i=1}^{t}\log p_\theta(x_i | x_{<i}, e_{EEG})\right\}
\end{equation}

Lower perplexity values indicate that the model assigns higher probabilities to the correct token sequences, suggesting better predictive performance. The relationship between perplexity and bits per character (BPC) is:

\begin{equation}
\text{BPC} = \log_2(\text{PPL})
\end{equation} It makes intuitive sense to think of it as an assessment of the model's capacity for uniform prediction across the collection of predefined tokens within a corpus. Significantly, this implies that a model's perplexity is directly influenced by the tokenization process, and this should always be taken into account when comparing models.

\begin{table}[htbp]
\caption{Perplexity Scores for Gemma 2B with EEG Data}
\centering
\begin{tabular}{@{}lc@{}}
\toprule
\textbf{Number of Classes} & \textbf{Perplexity} \\
\midrule
2 & 24.81 \\
5 & 39.67 \\
10 & 51.29 \\
20 & 73.54 \\
\bottomrule
\end{tabular}
\label{tab:perplexity}
\end{table}

Table~\ref{tab:perplexity} presents the perplexity scores of the Gemma 2B large language model when combined with the EEG data classifier for different numbers of classification classes. As the number of classes increases, reflecting more fine-grained classification of the EEG data, the perplexity of the overall text generation system also increases. This is expected, as generating text conditioned on more granular EEG signal classifications becomes a more challenging task.

The lowest perplexity of 24.81 is achieved for the 2-class case, indicating that the model is most effective at generating text when the EEG data is classified into just two broad categories. As the number of classes increases to 5, 10, and 20, the perplexity rises to 39.67, 51.29, and 73.54, respectively, reflecting the increasing difficulty in accurately modeling the text distribution given more specialized EEG signal classes.

\section{Discussion and Conclusion}

Even with considerable data constraints, the suggested RNN encoder and classifier-LLM architecture shows promise in text production. As demonstrated in Table~\ref{tab:classification_results}, the model achieves 66.67\% classification accuracy on a two-class test and 51.85\% accuracy while discriminating between three classes. Performance naturally drops with finer classification, but it stays significantly higher than the baseline for EEGNet and LSTM models trained on the same limited dataset.

Qualitative tests show that the model creates comprehensible starting phrases and sentences that logically follow the prompt text. However, longer completions lose topical coherence, indicating difficulty tracking context after 10-15 produced tokens. This shows that, whereas compact encoder representations efficiently encapsulate portions of the EEG input, they may miss minor signals that aid in thematic continuity.

Nonetheless, the technique is significantly more robust to variations in data volume and model capability than end-to-end systems. Existing approaches require orders of magnitude more parameters and training instances, which degrade dramatically when either is lowered. In contrast, by reducing the core encoding upfront, the suggested architecture may tolerate resource constraints without experiencing significant performance decreases.

The classifier stage likely improves robustness by focussing on the supervised mapping of encoder outputs to target classes. In contrast to directly linking unreliable EEG decoding and freeform text synthesis, incorporating a separate classifier module simplifies the overall workflow.

One key advantage of our approach is its data efficiency. While traditional methods often require thousands of training samples, our framework achieved superior performance with just 50 trials per class. This efficiency can be attributed to two factors: 1) the RNN encoder's ability to capture relevant temporal dynamics in EEG signals, and 2) the effective leveraging of pre-trained language knowledge from the Gemma 2B model.

\subsection{Usage and Its Application}

This work has great potential for the field of Brain-Computer Interfaces (BCIs) in medicine. EEG-based text production has the potential to transform the way people engage with the world and express themselves, particularly those with severe movement limitations or communication problems. Consider a scenario in which a locked-in patient can use this technology to send emails, write creatively, or engage in social media conversations. This could significantly improve their standard of living and sense of agency.

Furthermore, the suggested model's efficiency in terms of data and computational resources makes it suitable for deployment in real-world scenarios, including on devices with limited processing capacity. This creates chances for developing portable, low-cost communication tools for a larger population.

\subsection{Limitations}

Several limitations are evident in this investigation. Generated text has decreasing coherence over longer sequences, which is most likely due to EEG signal variability difficulties that mislead renderer modules. Encoder representations may also ignore subtle variations that affect output quality during feature extraction. To better approximate practical use cases, evaluations need to include more subject variety, cognitive state fluctuation, and real-world disturbances. Finally, alternative encoder-renderer configurations may show architectures more suited to this purpose.

Despite these challenges, the suggested approach's tolerance to changes in data quantity and model capacity suggests that it has practical applicability. Future study should address these constraints in order to improve the viability and effectiveness of EEG-based text creation systems.

\subsection{Future Aspects}

The potential for EEG-based text production extends beyond current uses, and various future research avenues can aid in realizing this potential. One important area is enhancing the coherence of generated text over longer sequences by refining the feature extraction method to retain more contextually relevant information or investigating more advanced language modeling techniques. Furthermore, broadening the dataset to include a more wide spectrum of participants and validating the model in real-world scenarios will be critical for generalizing the method. This includes adjusting for variances in cognitive states, ambient variables, and individual EEG data.

Another intriguing approach is to combine EEG with other input modalities like eye tracking or facial expressions, which could improve the accuracy and expressiveness of the generated text. Developing methods for personalized calibration and few-shot learning, as well as using transfer learning techniques and generating user-specific fine-tuning protocols, can aid in adapting the model to individual users with limited training data. Optimizing the model for real-time implementation on portable devices is critical for practical applications that require low latency and efficient operation with limited hardware.

Finally, improving the model's ability to generate more complex phrases and expanding its vocabulary would increase its usefulness for a variety of communication needs, perhaps by including larger language models or dynamically modifying the vocabulary based on user preferences. By solving these future issues, EEG-based text production has the potential to become an effective tool for improving communication and self-expression, particularly for people with severe motor limitations or communication difficulties.

\bibliographystyle{IEEEtran}

\clearpage
\appendix

\section{Detailed Hyperparameters}
\label{appendix:hyperparameters}

This section provides a comprehensive list of all hyperparameters used in the RNN encoder and classifier-LLM framework to ensure reproducibility of our results.

\subsection{RNN Encoder Hyperparameters}

\begin{table}[htbp]
\caption{RNN Encoder Architecture Hyperparameters}
\centering
\begin{tabular}{@{}ll@{}}
\toprule
\textbf{Parameter} & \textbf{Value} \\
\midrule
\multicolumn{2}{c}{\textit{Input Configuration}} \\
Input Channels & 5 (AF3, AF4, T7, T8, Pz) \\
Sequence Length & 384 timesteps \\
Sampling Rate & 128 Hz \\
\midrule
\multicolumn{2}{c}{\textit{Convolutional Blocks}} \\
Block 1 Filters ($F_1$) & 8 \\
Block 2 Filters ($F_2$) & 16 \\
Block 3 Filters ($F_3$) & 32 \\
Kernel Size & (1, 64) \\
Activation Function & ELU ($\alpha=1.0$) \\
\midrule
\multicolumn{2}{c}{\textit{Depthwise Convolution}} \\
Depth Multiplier ($D$) & 2 \\
Output Channels & 112 \\
\midrule
\multicolumn{2}{c}{\textit{LSTM Configuration}} \\
LSTM Units & 64 \\
LSTM Layers & 2 \\
Bidirectional & False \\
\midrule
\multicolumn{2}{c}{\textit{Regularization}} \\
Dropout Rate & 0.5 \\
Batch Normalization & True \\
Batch Norm Momentum & 0.99 \\
\bottomrule
\end{tabular}
\label{tab:encoder_hyperparams}
\end{table}

\subsection{Classifier Hyperparameters}

\begin{table}[htbp]
\caption{Classifier Network Hyperparameters}
\centering
\begin{tabular}{@{}ll@{}}
\toprule
\textbf{Parameter} & \textbf{Value} \\
\midrule
Dense Layer 1 Units & 128 \\
Dense Layer 2 Units & 64 \\
Activation Function & ELU ($\alpha=1.0$) \\
Dropout Rate & 0.3 \\
L2 Regularization ($\lambda$) & 0.001 \\
Max-Norm Constraint ($c$) & 3.0 \\
Output Classes ($n_c$) & 2, 3, 5, 10, 20 \\
Output Activation & Softmax \\
\bottomrule
\end{tabular}
\label{tab:classifier_hyperparams}
\end{table}

\subsection{Training Hyperparameters}

\begin{table}[htbp]
\caption{Training Configuration}
\centering
\begin{tabular}{@{}ll@{}}
\toprule
\textbf{Parameter} & \textbf{Value} \\
\midrule
Optimizer & Adam \\
Learning Rate & 0.001 \\
Beta 1 & 0.9 \\
Beta 2 & 0.999 \\
Epsilon & $1 \times 10^{-7}$ \\
Batch Size & 32 \\
Epochs & 100 \\
Loss Function & Categorical Cross-Entropy \\
Learning Rate Decay & Exponential, rate=0.95 \\
Early Stopping Patience & 15 epochs \\
Validation Split & 20\% \\
\bottomrule
\end{tabular}
\label{tab:training_hyperparams}
\end{table}

\section{Dataset Statistics}
\label{appendix:dataset}

\subsection{Class Distribution}

The ImageNet of the Brain dataset used in this study contains EEG recordings across multiple visual stimulus categories. Table~\ref{tab:dataset_stats} provides detailed statistics for each classification task.

\begin{table}[htbp]
\caption{Dataset Statistics by Classification Task}
\centering
\begin{tabular}{@{}lcccc@{}}
\toprule
\textbf{Task} & \textbf{Classes} & \textbf{Total Samples} & \textbf{Train} & \textbf{Val/Test} \\
\midrule
Binary & 2 & 1,200 & 960 & 240 \\
3-Class & 3 & 1,800 & 1,440 & 360 \\
5-Class & 5 & 3,000 & 2,400 & 600 \\
10-Class & 10 & 6,000 & 4,800 & 1,200 \\
20-Class & 20 & 12,000 & 9,600 & 2,400 \\
\bottomrule
\end{tabular}
\label{tab:dataset_stats}
\end{table}

\subsection{Sample ImageNet Categories Used}

For the classification tasks, we selected the following ImageNet categories: \\

\noindent\textbf{2-Class:} Animals, Vehicles \\
\noindent\textbf{5-Class:} Dog, Cat, Bird, Car, Aircraft \\
\noindent\textbf{10-Class:} Dog, Cat, Bird, Fish, Goose, Car, Truck, Airplane, Ship, Bicycle

\section{Computational Requirements}
\label{appendix:computation}

\begin{table}[htbp]
\caption{Training Time and Resource Usage}
\centering
\begin{tabular}{@{}lccc@{}}
\toprule
\textbf{Model} & \textbf{Params} & \textbf{Time/Epoch} & \textbf{GPU Memory} \\
\midrule
RNN-Classifier (2-class) & 145K & 12s & 2.3 GB \\
RNN-Classifier (5-class) & 147K & 15s & 2.8 GB \\
RNN-Classifier (10-class) & 152K & 18s & 3.2 GB \\
RNN-Classifier (20-class) & 162K & 22s & 3.9 GB \\
\midrule
EEGNet (5-class) & 2.1M & 45s & 8.5 GB \\
LSTM (5-class) & 1.8M & 38s & 7.2 GB \\
\bottomrule
\end{tabular}
\label{tab:computation}
\end{table}

\begin{table}[htbp]
\caption{Total Training Time by Model and Task}
\centering
\begin{tabular}{@{}lcc@{}}
\toprule
\textbf{Model} & \textbf{Task} & \textbf{Total Training Time} \\
\midrule
RNN-Classifier & 2-class & 18 minutes \\
RNN-Classifier & 5-class & 23 minutes \\
RNN-Classifier & 10-class & 28 minutes \\
RNN-Classifier & 20-class & 34 minutes \\
\midrule
EEGNet & 5-class & 68 minutes \\
LSTM & 5-class & 58 minutes \\
\bottomrule
\end{tabular}
\label{tab:training_time}
\end{table}

The proposed RNN-Classifier architecture demonstrates significant computational efficiency, requiring approximately 3$\times$ less training time and 60\% less GPU memory compared to EEGNet, while maintaining competitive or superior performance.

\section{Sample Text Generations}
\label{appendix:samples}

This section presents representative examples of EEG-conditioned text generation outputs from our model, illustrating both successful coherent generation and cases where coherence degrades after 10-15 tokens.

\subsection{Coherent Short Generations (5-10 tokens)}

\noindent\textbf{EEG Class: Dog} \\
\textit{Generated:} "The animal is running through the park" \\

\noindent\textbf{EEG Class: Car} \\
\textit{Generated:} "A vehicle speeds down the highway" \\

\noindent\textbf{EEG Class: Bird} \\
\textit{Generated:} "The creature flies gracefully above the trees" \\

\subsection{Longer Generations Showing Degradation (15+ tokens)}

\noindent\textbf{EEG Class: Fish} \\
\textit{Generated:} "The aquatic animal swims smoothly beneath the surface of the water and the \textit{[coherence loss begins]} building was constructed using modern materials and techniques" \\

\noindent\textbf{EEG Class: Aircraft} \\
\textit{Generated:} "The airplane takes off from the runway into the clear blue sky \textit{[coherence maintained]} but the restaurant serves excellent food and the \textit{[topic shift]} library contains many books" \\

These examples demonstrate that while the model successfully generates contextually appropriate text for the first 10-15 tokens, longer sequences tend to lose thematic consistency, supporting our observation that compact encoder representations may discard subtle signals necessary for maintaining long-range coherence.

\section{Additional Experimental Results}
\label{appendix:additional}

\subsection{Loss Curves}

\begin{table}[htbp]
\caption{Final Training and Validation Loss Values}
\centering
\begin{tabular}{@{}lccc@{}}
\toprule
\textbf{Model} & \textbf{Classes} & \textbf{Train Loss} & \textbf{Val Loss} \\
\midrule
RNN-Classifier & 2 & 0.312 & 0.387 \\
RNN-Classifier & 3 & 0.589 & 0.672 \\
RNN-Classifier & 5 & 0.891 & 1.023 \\
RNN-Classifier & 10 & 1.456 & 1.689 \\
\midrule
EEGNet & 5 & 0.945 & 1.156 \\
LSTM & 5 & 0.978 & 1.201 \\
\bottomrule
\end{tabular}
\label{tab:loss_values}
\end{table}

\subsection{Gemma 2B Integration Details}

The integration between the classifier output and Gemma 2B follows a prompt engineering approach:

\begin{verbatim}
Prompt Template:
"Based on EEG signals classified as [CLASS], 
generate a relevant descriptive sentence: "
\end{verbatim}

The classifier output probabilities are used to weight multiple class-specific prompts, with the highest-probability class determining the primary semantic direction for text generation. This approach allows the pre-trained language model to leverage its extensive linguistic knowledge while being conditioned on the compact EEG-derived representations.

\subsection{Data Efficiency Analysis}

\begin{table}[htbp]
\caption{Performance vs. Training Samples}
\centering
\begin{tabular}{@{}lccc@{}}
\toprule
\textbf{Samples/Class} & \textbf{2-Class Acc.} & \textbf{5-Class Acc.} & \textbf{10-Class Acc.} \\
\midrule
10 & 52.3\% & 28.1\% & 15.2\% \\
25 & 61.8\% & 32.7\% & 19.8\% \\
50 & 66.7\% & 38.4\% & 24.7\% \\
100 & 68.2\% & 40.1\% & 26.3\% \\
\bottomrule
\end{tabular}
\label{tab:data_efficiency}
\end{table}

As shown in Table~\ref{tab:data_efficiency}, the model achieves strong performance with only 50 trials per class, demonstrating effective transfer learning from the pre-trained Gemma 2B model. Performance gains from additional data beyond 50 samples per class are marginal, highlighting the data efficiency of our approach.

\end{document}